# Library of Congress Subject Heading (LCSH) Browsing and Natural Language Searching

**Charles-Antoine Julien**
School of Information Studies
McGill University, Montreal, Canada
charles.julien@mcgill.ca

**Banafsheh Asadi, Jesse D. Dinneen, Fei Shu**
School of Information Studies
McGill University, Montreal, Canada
banafsheh.asadi@mail.mcgill.ca;
jesse.dinneen@mail.mcgill.ca;
fei.shu@mail.mcgill.ca

## ABSTRACT

Controlled topical vocabularies (CVs) are built into information systems to aid browsing and retrieval of items that may be unfamiliar, but it is unclear how this feature should be integrated with standard keyword searching. Few systems or scholarly prototypes have attempted this, and none have used the most widely used CV, the Library of Congress Subject Headings (LCSH), which organizes monograph collections in academic libraries throughout the world. This paper describes a working prototype of a Web application that concurrently allows topic exploration using an outline tree view of the LCSH hierarchy and natural language keyword searching of a real-world Science and Engineering bibliographic collection. Pilot testing shows the system is functional, and work to fit the complex LCSH structure into a usable hierarchy is ongoing. This study contributes to knowledge of the practical design decisions required when developing linked interactions between topical hierarchy browsing and natural language searching, which promise to facilitate information discovery and exploration.

## Keywords

Information Exploration Systems, Information Retrieval, Information Organization, Information Visualization.

## INTRODUCTION

To successfully use current searching tools users must generally possess or acquire the vocabulary used by the authors of the relevant documents. Searchers inevitably encounter information needs that they cannot adequately express with one or two vague and broad keywords that often have multiple meanings. This *vocabulary mismatch problem* plagues information retrieval systems — especially when searching unfamiliar topical areas. This paper is a design case study of an online information exploration tool designed to address this problem by capitalizing on existing information organization investments using the Library of Congress Subject Headings (LCSH) to complement natural language information retrieval in a real-world bibliographic collection from the domain of science and engineering.

## Issues with Natural Language Searching

When facing unsatisfactory results in unstructured text collections, searchers must formulate multiple queries and consider long lists of often irrelevant items in the hopes of stumbling upon adequate search vocabulary; this can be frustrating and time consuming (Dork, Williamson, & Carpendale, 2012), and users' reliance on results ranking biases their selections toward the top of the first ten results regardless of the actual quality of the results. There is a growing interest towards supporting more exploratory search, especially when searchers are seeking information in unfamiliar domains where they may not initially possess adequate search vocabulary.

## Controlled Topical Vocabulary (CV)

CV is partly designed for searchers to explore topical groups of information. CV can be useful and appreciated (Hearst, 2006) when searchers perform search tasks in unfamiliar domains that require multiple results in order to gain a broader understanding of the topic. Manually created categories and their relations (i.e., topic models or ontologies) are preferred by users over those created by automated clustering methods.

Maintained by the Library of Congress (LC) since 1898, LCSH has been continuously evolving for the creation of topical access points and their assignment to bibliographic records. Despite perennial criticisms centered on cost, scalability and consistency, LCSH is the standard list of CV for most libraries in the United States and throughout the world. LCSH is therefore a promising case study of a widely implemented, large, and mature general knowledge topic model.

Current LCSH browsing systems (see http://authorities.loc.gov/) do not provide explicit features to navigate the LCSH structure, forcing users to visit each LCSH individually and the latest library faceted Discovery tools also ignore relations between LCSH topics. Academic LCSH browsing prototypes (Yi & Chan, 2010) are designed as interfaces to the empty LCSH structure from LC that does not contain a collection. Like a small number of previous studies (Chan, Wu, Talbot, Cammarano, & Hanrahan, 2008; Dork et al., 2012), this research strives to seamlessly integrate both searching and browsing functionality; this study uniquely uses the LCSH topic network and asks the following research questions:

RQ1. How can we visually represent a collection's LCSH topic structure using recognized interactions?

RQ2. How should keyword searching affect the interactive LCSH structure?

RQ3. How should interactions with the LCSH structure change the results of keyword searching?

Answers to these questions will partly address the lack of knowledge concerning how best to integrate CV exploration and natural language searching.

## METHODOLOGY
The following section describes the datasets that were computationally treated to extract and simplify the LCSH hierarchy that represents the collection. Technology choices and the prototype are described followed by issues encountered.

### Datasets
A collection organized using LCSH is generally comprised of two data sources: 1) bibliographic records that are indexed by topic with at least one topical LCSH string, and 2) the LCSH *authority* records that contain the *established* topical LCSH strings and their broader topics, which define the topical LCSH hierarchy. Our dataset is comprised of 1) a collection of 122,197 bibliographic records housed at McGill University's Schulich Library of Science and Engineering, and 2) its roughly 205,000 LCSH authority records. The domain of science was selected as an extreme test for this study since it offers the deepest and most connected portion of the LCSH structure (Julien, Tirilly, Leide, & Guastavino, 2012; Yi & Chan, 2010).

### Making a Usable LCSH Hierarchy
Representing a collections' LCSH structure as a hierarchy is a promising approach for improving topical exploration since it permits its visual representation and interaction using universally recognized outline views (e.g., computer file explorer) that allow systematic top-down navigation (also called *drill-down*) from broad to specific topics. LCSH authority records contain broader/narrower term relationships but it was not conceived as a hierarchical thesaurus; this is addressed computationally based on Julien et al. (2012) who have retrofitted LCSH into a topical hierarchy. The process extracts the LCSHs assigned to the bibliographic collection, which are matched to an authority record to place each bibliographic record in the LCSH hierarchy.

The transformation of the LCSH structure into a hierarchy highlights the issue of *multiple inheritances* found in LCSH and other CVs such as Medical Subject Headings (MeSH). This refers to topics having multiple immediate broader terms (i.e., parent topics) that are difficult to represent in a topical hierarchy. This issue is usually addressed by duplicating narrower terms, and all their descendants if any, under all their parents. This can have a significant impact on the size of the tree since duplicating a broad subject such as Engineering creates copies of its hundreds of narrower term descendants. Applied to our datasets, this solution produces a highly redundant LCSH hierarchy of 2.5 million nodes where 99.2% of topics have at least two occurrences in the tree. Users may naturally recognize topics they have previously visited under a different branch; therefore, the initial version of the Web application would not explicitly communicate that most topics have at least one duplicate.

Some researchers have developed algorithms to modify large topic structures that can overwhelm users; our LCSH hierarchy was simplified using the automated process described in Julien et al. (2013). The approach opportunistically prunes a topic tree based on the expected power law distribution of topic assignments to the collection, which are common in organized information (Julien et al., 2012). Capitalizing on this trend, Julien et al. (2013) reduce the size and depth of a collection's subject tree by pruning the subjects that are not representative of a collection, which was applied to our LCSH hierarchy. The implemented LCSH hierarchy contains 8,760 unique topics whose multiple inheritance duplicates produce a tree containing 578,449 topics.

### Human-Information Interface
The system is a two panel coordinated multiple view system (CMV) where user's interaction in one view automatically changes the other. There is currently no CMV system for library collections and its LCSH topic hierarchy. The initial prototype shows: 1) a traditional outline tree view of the LCSH hierarchy and 2) the ranked list of results from natural language searching of the collection. The coordination of these two panels currently entails that keyword search results are ranked and listed as usual, results are positioned in the LCSH tree to conspicuously reveal promising topical branches to explore, and selecting topics in the hierarchy can act as a filter in conjunction with or instead of natural language searching of the collection.

### Development Technologies
Integrating the LCSH hierarchy and keyword search features entailed considerable technical and design challenges over 20 months of development. The system queries a MySQL database and Solr search index using Java and the Vaadin Web development framework.

## RESULTS

Figure 1 shows a left panel that contains the LCSH hierarchy, and a right panel that, depending on the user's last action, either presents the bibliographic records assigned to the selected LCSH or the ranked results of a natural language keyword search. Figure 1 shows an example where the user has performed a natural language search for the keyword "finite element", matching bibliographic records are ranked in the right panel, while the most promising branches of the LCSH hierarchy are indicated using green arrows. The current definition of *most promising* are the top two most assigned topics within the first 100 ranked results. Alternatively, each bibliographic record shown in the right panel lists its assigned topics as hyperlinks that users select to expand the topic branch in the LCSH hierarchy.

### Implications of Multiple Inheritance

Given that the vast majority of LCSHs have copies in multiple branches of the LCSH tree, 1) it is not obvious which copy should be highlighted in the tree when the user selects a hyperlinked topic from the ranked list of results, or

2) to which copy of a promising topic users should be guided after a keyword search. There is a lack of system design literature concerning how to design interactions with CV hierarchies that have multiple inheritances. The current version of the system uses a heuristic approach to target a specific copy of a topic: it will favor a topic copy whose branch shares the most nodes with the last selected topic, which leads users to the topic copy that is closest to their current location in the LCSH hierarchy.

## PILOT TESTING

Testing was done using a list of 28 searching and topic browsing tasks used by Julien (2010) for this collection. These covered tree topology navigation (e.g., find a common ancestor; find a specific topic at various depths), simple retrieval tasks whose wording contains valid search terms (e.g., Look for information on climate change), and complex retrieval tasks whose wording does not contain valid search terms for the collection (e.g., How would you fix a leaky faucet?). The student tester had never used the prototype system but was able to complete the tasks in about an hour. This interaction revealed that the interface is functional but it may have to communicate that a topic branch has copies elsewhere in the tree, one of which may have already been explored. Although identifying previously visited topics is not difficult, it is not known if searchers would understand why they meet the same topics under multiple topical contexts. The next iteration of the system will attempt to visually identify previously visited topics using a conspicuous color.

**Figure 1. Prototype search and topic exploration interface where users can explore the LCSH topic hierarchy (left) that contains the collection, which can also be searched using keywords(right).**

## ANALYSIS
RQ1 asked if the LCSH topic structured could be visually represented using recognized functionality: the system developed for this study demonstrates it is possible to fit LCSH in a usable yet highly redundant topic tree navigated using a traditional outline tree view.

RQ2 asked how keyword searching should affect the LCSH hierarchy: the system guides the users towards the two topic branches that contain the most search results. This places most results in the topic hierarchy where users may learn new search terms while gleaning related topics. Other implementations are possible but this one is a kind of search directed navigation (Fitchett, Cockburn, & Gutwin, 2014) where users are visually guided to promising browsing areas based on keyword search results.

RQ3 asked how interactions with the LCSH structure should change the results of keyword search: guided by current library catalogue faceted discovery interfaces, the prototype shows the bibliographic records that contain both the keywords and the selected topic. Other approaches are possible; for example, selecting a hierarchy topic might show search results assigned to any descendant of that topic, which would have a significant impact for a broad topic such as *Cybernetics* or *Engineering*.

### Multiple Inheritances
The issue of multiple inheritances is found in other CVs such as MeSH; therefore, advancing design knowledge concerning interactions with multiple inheritance CVs is likely beneficial to information exploration in many CV networks. When browsing a highly redundant topic hierarchy users can find a single topic via different routes. This may be desirable when searching for one of these duplicated subjects; however, it may be detrimental if a user is searching for a single non-duplicated topic buried within a very large and redundant hierarchy. There are no obvious preferred values of topic hierarchy redundancy and its proper optimization for a usable topic tree is an open question.

## LIMITATIONS
This study may not be generalizable to knowledge domains beyond science and engineering whose topic network might differ significantly (e.g., collections from the humanities).

## CONCLUSION
This research addresses the issue that organized collections have difficulty demonstrating the value created by expensive manual indexing using CVs that are not often used explicitly or understood by its users. The developed system allows users to explore an LCSH topic hierarchy in conjunction with natural language keyword searching: it is the only system that fits the large and structurally complex LCSH network into a usable outlined tree view in coordination with keyword search of the organized collection. As such, the system shows the promise of developing usable interface solutions to navigate CV networks with multiple inheritances in conjunction with traditional information retrieval. Ongoing development efforts are focused on managing the computational needs of the large redundant LCSH tree and ensuring its usability. This will be followed by usability tests and retrieval performance comparisons with traditional, separated browsing and searching systems.

## ACKNOWLEDGEMENTS
This research is supported by the National Science and Engineering Research Council of Canada (NSERC), the Fonds de recherche du Québec – Société et culture (FRQSC), and in kind contributions from McGill libraries.

# Library of Congress Subject Heading (LCSH) Browsing and Natural Language Searching

**C-A. Julien, B. Asadi, J. D. Dinneen, F. Shu**

School of Information Studies, McGill University, Montreal, Canada

charles.julien@mcgill.ca

## What

Our Web app explores integrating topic exploration with searching.

**RQ1.** How can we visually represent a collection's LCSH topic structure using recognized interactions?

**RQ2.** How should keyword searching affect the interactive LCSH structure?

**RQ3.** How should interactions with the LCSH structure change the results of keyword searching?

## Why

- Some information needs cannot be expressed with a few broad keywords, and unfamiliar domain vocabulary worsens this.
- If linked, topical hierarchy browsing and natural language searching could facilitate information discovery and exploration but this has never been done.
- We contribute to the practical design decisions required when integrating and developing interactions between the two.

## How

A design case study of an online information exploration tool using a science and engineering collection at McGill.

**Conclusion:**

- The developed system allows users to explore an LCSH topic hierarchy in conjunction with natural language keyword searching.
- The system shows the promise of developing usable interface solutions to navigate CV networks with multiple inheritances in conjunction with traditional information retrieval.

### Browse

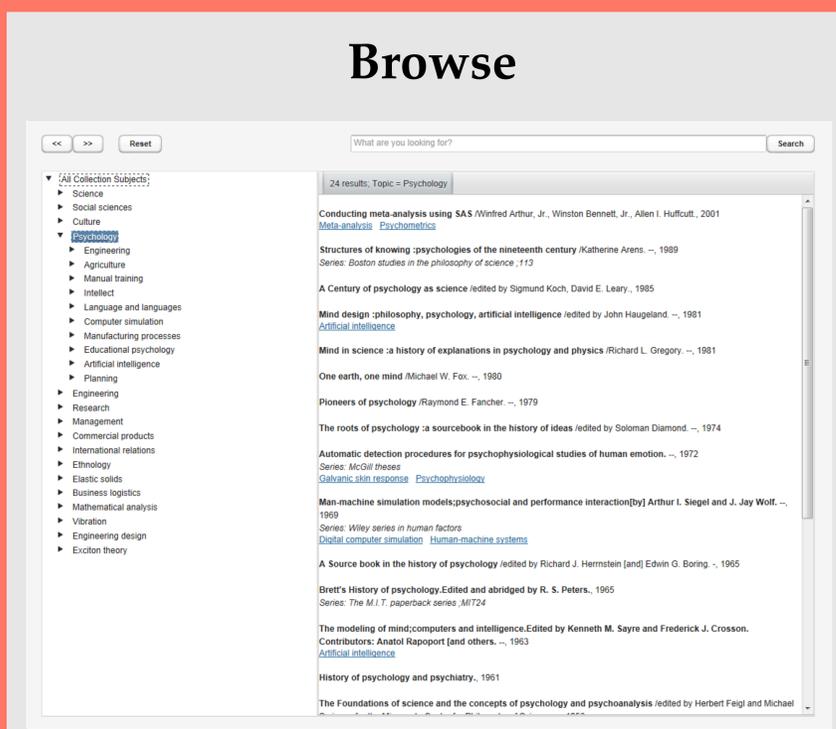

### Integration

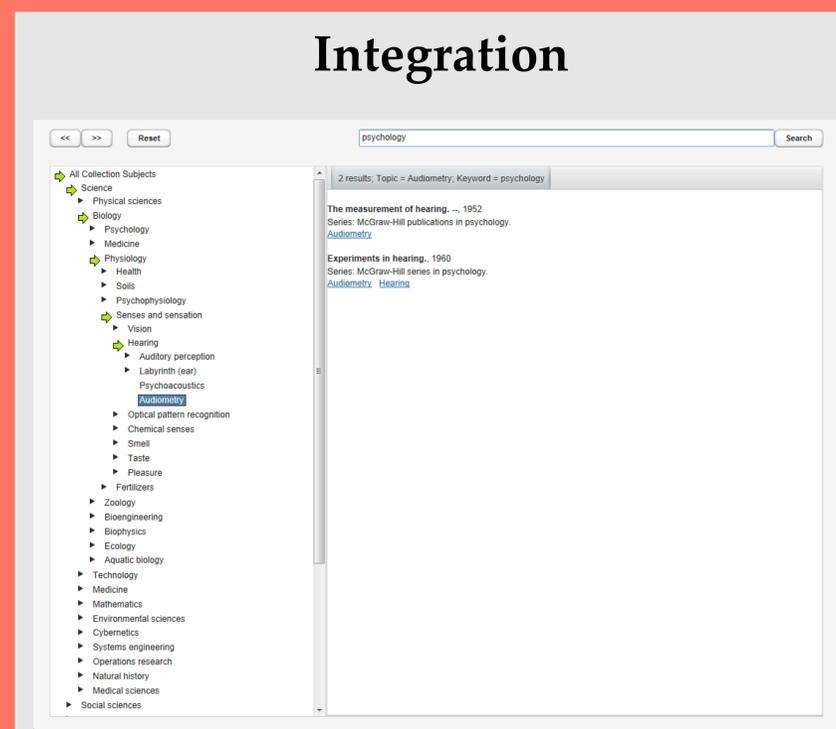

### Search

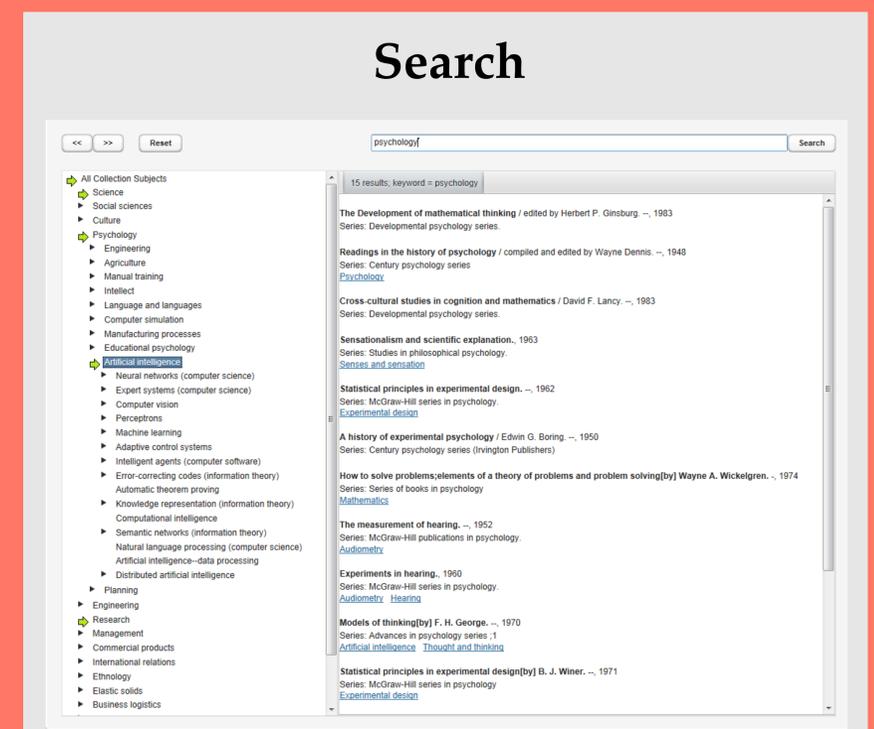